# KRAS G12D protein screening for pancreatic cancer clinical trials using an AlGaN/GaN high electron mobility transistor biosensor

Sheng-Ting Hung[1], Cheng Yan Lee[1], Chen-Yu Lien[1], Cheng-Hsuan Chan[1], Ya-Han Yang[2], Quark Yungsung Chen[3,4], Kuang-Hung Cheng[5], Kung-Kai Kuo[2,6*], Li-Wei Tu[1*], Ching-Wen Chang[7*]

[1]Department of Physics, National Sun Yat-sen University, Kaohsiung 80424, Taiwan

[2]Department of Surgery, Kaohsiung Medical University Hospital, Kaohsiung 80756, Taiwan

[3]Industry Academia Innovation School, National Yang Ming Chiao Tung University, Hsinchu 300093, Taiwan

[4]Department of Physics and Texas Center for Superconductivity, University of Houston, Houston, TX 77204 USA

[5]Institute of Biomedical Sciences, National Sun Yat-sen University, Kaohsiung 80424, Taiwan

[6]Department of Surgery, E-DA Healthcare Group E-DA Dachang Hospital, Kaohsiung 80785, Taiwan

[7]Institute and Undergraduate Program of Electro-Optical Engineering, National Taiwan Normal University, Taipei 116059, Taiwan

**Abstract**

Clinical trials screening KRAS G12D protein for 30 pancreatic ductal adenocarcinoma (PDAC) patients and 30 healthy donors were conducted utilizing an AlGaN/GaN high electron mobility transistor (HEMT) biosensor. All resistance change ratios of PDAC patients are higher than the standard deviation above the mean resistance change ratio obtained from all healthy donors. The results demonstrate the effectiveness of the HEMT biosensor and reveal its potential for early detection of pancreatic cancer with KRAS G12D protein screening.

**1. Introduction**

Pancreatic cancer is highly lethal, with a five-year survival rate of 12.8% from 2014 to 2020[1]. Pancreatic cancer is difficult to diagnose at an early stage as it often has subtle or no symptoms. As a result, it is usually diagnosed at an advanced stage when treatment options are limited and less effective. The lack of effective early detection methods is currently a global bottleneck in the treatment of pancreatic cancer. Therefore, developing early detection methods with high specificity and sensitivity is crucial to improving the survival rates of patients with pancreatic cancer.

The Rat sarcoma (RAS) gene family, including Kirsten RAS (KRAS), neuroblastoma RAS (NRAS), and Harvey RAS (HRAS), represents the most commonly mutated oncogenes in human cancers. Among the RAS family, KRAS is the most frequently mutated RAS gene in cancers, accounting for 84% of all RAS missense mutations, followed by NRAS at 12%, while HRAS mutations are rare, occurring in only 4% of cases[2]. In addition, KRAS is mutationally activated in 94% of pancreatic ductal adenocarcinoma (PDAC)[2]. Among all KRAS mutations, KRAS G12D mutation is a highly specific biomarker for pancreatic cancer, particularly PDAC[3]. This mutation is found in a significant majority of PDAC cases[2], making it a potential critical target for diagnosis and treatment. Being one of the earliest genetic alterations in PDAC development, the presence of the KRAS G12D mutation can help in the early detection and diagnosis of pancreatic cancer[4, 5]. Although targeting KRAS G12D mutant has been challenging, recent advances have led to the development of potential inhibitors and immunotherapeutic approaches[6]. Moreover, the detection of KRAS mutations has also drawn great research interest. Current methods for detecting the KRAS G12D mutation include a variety of advanced molecular detection techniques. Digital droplet PCR (ddPCR) and next-generation sequencing (NGS) are often used for their high sensitivity and specificity in identifying KRAS mutations[7]. Additionally, Taqman allele-specific qPCR is employed for multiplex detection of KRAS mutations[8]. A multiplex detection for 3 DNA mutations based on polymerase chain reaction (PCR) and surface enhanced Raman spectroscopy (SERS) was demonstrated with plasma specimens[9]. A liquid biopsy-based early diagnosis of pancreatic cancer precursors was conducted based on single molecule with a large transistor (SiMoT) technique[10]. Though these techniques demonstrated their potential for early detection of KRAS mutations, the detection of DNA may be technically demanding and require sophisticated equipment. In addition, DNA samples are susceptible to contamination, which may lead to false results. In contrast, antigen-antibody detection is relatively simple yet offers high specificity and sensitivity. The fact that oxidative stress results in KRAS G12D protein release from PDAC cells[11, 12] implies the potential of detecting KRAS G12D proteins from blood specimens of PDAC patients.

High Electron Mobility Transistor (HEMT) electronic devices provide sensitive electrical responses to biomarker detection owing to their high electron mobility of the two-dimensional electron gas (2DEG). In particular, the high density and mobility of the 2DEG at the AlGaN/GaN heterojunction interface, induced by polarization effects, enhance the sensitivity of their biosensing applications. Recent advancements have demonstrated the capability of AlGaN/GaN HEMT-based biosensors to detect specific biomolecules with high sensitivity and specificity[13]. We have previously reported a time-effective sensing platform for the rapid detection of the carbohydrate antigen 19-9 (CA 19-9) and standard carcinoembryonic antigen (CEA) tumor markers by leveraging the high sensitivity of the AlGaN/GaN HEMT biosensing chip[14, 15]. We obtained 94% accuracy rate in the previous clinical trials using serums from 35 pancreatic cancer patients[15]. Though CA 19-9 is a validated tumor biomarker for pancreatic cancer prognosis, its effectiveness remains controversial[16-21]. Thus, we further extend our study to detecting KRAS G12D protein. The isoelectric point of KRAS G12D protein is 6.33 according to PhosphoSitePlus Database[22]. Therefore, the protein is negatively charged at PH 7.4 phosphate buffered saline (PBS). When KRAS G12D proteins attach to the surface of the HEMT sensing area, the negative charges repel the 2DEG resulting in an increase in resistance. Herein we report the results of clinical trials of 30 PDAC patients and 30 health donors on KRAS G12D protein detection using our HEMT biosensing chip.

## 2. Materials and Methods

*2.1 Device fabrication and measurement setups*

AlGaN/GaN HEMT wafers were provided by Unikorn Semiconductor Corporation. The structure of HEMT wafers and the device fabrication and surface modification were described in our previous work[15], except that we substituted the CA 19-9 antibody with the RAS G12D antibody purchased from Invitrogen. The electrical setup was described in our previous work[15] and the source-drain resistance ($R_{sd}$) was recorded over time using a precision resistance meter (Hioki, RM3544-01) during the measurement. The fabrication of the electrodes was similar to the photolithography procedure we previously reported[15], with an extra coating of $SiO_2$ to protect the surface of the HEMT chip and replacing the Ti/Al/Ti/Au (20 nm/160 nm/20 nm/50 nm) with Ti/Al/Ti/Al (20 nm/160 nm/20 nm/80 nm) layers to reduce the cost.

The device fabrication and surface functionalization are briefed below. A layer of 200 nm thick $SiO_2$ was coated on the surface of the chip using high-density plasma chemical vapor deposition (HDP-CVP, Advances System Technology Co., Ciede-200). Afterward, positive photoresist AZ-1500 was spin-coated on a clean substrate and exposed to ultraviolet (UV) light with a customized mask aligned by a mask aligner. The exposed areas designed for electrodes were removed by the developer, followed by cleaning with deionized (DI) water and drying with nitrogen. The dried chip was treated with Inductively Coupled Plasma (ICP) etching to remove the $SiO_2$ coating on the electrodes, then underwent the e-beam evaporation of Ti/Al/Ti/Al (20 nm/160 nm/20 nm/80 nm) layers followed by a lift-off process. The ohmic contacts of the electrodes were achieved by rapid thermal annealing (RTA). The low work function of TiN formed on the interface between nitride and metal alloy after annealing and the spiking mechanism led to good contact of TiN to the 2DEG[23], implying energy-efficient electron transport. Subsequently, the chip was again coated with a layer of 200 nm thick $SiO_2$ to protect the electrodes from damage caused by the surface modification on the sensing area later. The positive photoresist AZ-1500 was spin-coated on top of the $SiO_2$ layer followed by the photolithography procedure to allow only the sensing area and prob contact areas to expose the $SiO_2$ layer. The $SiO_2$ layer on the sensing area and prob contact areas was removed by immersing the chip in the Buffered Oxide Etch (BOE) to expose the HEMT surface. The remaining photoresist was then removed by sequentially rinsing with acetone, isopropanol and DI water. The sensing area of the HEMT surface was subsequently functionalized with 3-Aminopropyltrimethoxysilane (APTMS) and stored. Before measuring samples, 10 $\mu$L of 1% RAS G12D antibody was deposited onto the sensing area for 30 minutes, then the solution was removed from the sensing area and the same step was repeated one time.

2.2 Preparation of standard KRAS G12D solution

KRAS G12D protein was purchased from SignalChem Biotech Inc. To determine the detection limit of the HEMT biomedical sensor, KRAS G12D protein was diluted to a series of concentrations with the buffer solution. The buffer solution was prepared by mixing phosphate-buffered saline (PBS) with 2% bovine serum albumin (BSA).

2.3 Clinical specimens

Human serum specimens from 30 patients and 30 health donors were collected from Kaohsiung Medical University following relevant ethical requirements. Specimens were collected from volunteers excluding those with blisters, inflammation, blood-borne infectious diseases, tuberculosis, diabetes, cardiovascular disease, peptic ulcers, hypertension, kidney disease, asthma, colds, acute infections, infectious diseases, or allergic conditions.

Five milliliters of whole blood samples were collected and then allowed to clot by leaving it undisturbed at room temperature for a few minutes. The clot was removed by centrifuging at 2,500 rpm for 10 minutes at 4 °C. The resulting serum specimens were stored at -80 °C for further use to avoid component deterioration.

2.4 Resistance measurement

A drop of 10 $\mu$L buffer solution was first deposited on the sensing area of the HEMT biosensing chip to measure the background resistance of the chip. After measuring the buffer solution, the buffer was removed from the sensing area, followed by adding 10 $\mu$L of standard KRAS G12D samples or 10 $\mu$L of clinical specimens. For measuring a series of KRAS G12D proteins with various concentrations, standard KRAS G12D solutions with increasing concentrations were deposited on the sensing area after removing the buffer solution or previous sample with a lower concentration of KRAS G12D protein.

**Results and Discussion**

*3.1 Functionalization of HEMT biosensor*

A set of typical $I_{sd}$-$V_{sd}$ curves before and after surface modifications of the HEMT biosensor is plotted in Figure 1. The results are similar to the previously reported HEMT biosensors and the causes of such variations were due to the influence of net charges on the surface of the sensing area above the 2DEG after surface modifications.

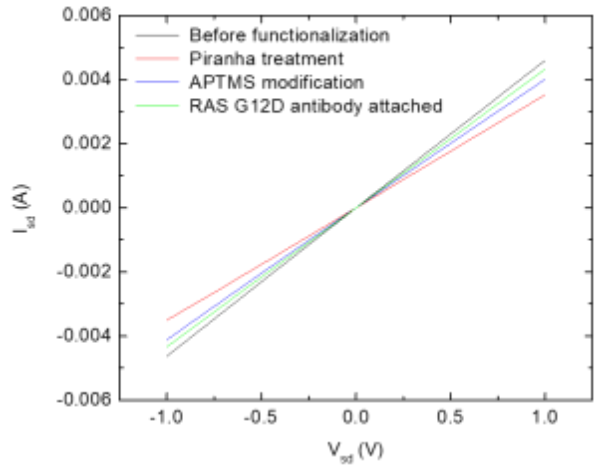

Figure 1. $I_{sd} - V_{sd}$ curves measured before and after surface modifications of the HEMT biosensor.

*3.2 Measurement of standard KRAS G12D solution*

KRAS G12D proteins with various concentrations were measured to quantify the detection limit of this biosensing chip. A typical result of such measurement is shown in Figure 2(a). We define the rate of change (RC) in resistance as

$$RC = \frac{\Delta \bar{R}}{\bar{R}} \times 100\%,$$

where $\bar{R}$ is the average resistance value measured from one drop of sample, and $\Delta \bar{R}$ is the difference of $\bar{R}$ values obtained in a sequential measurement, i.e. to take an $\bar{R}$ subtracting the former $\bar{R}$ value. The rate of change results from 3~5 sets of measurements were averaged and plotted as a function of KRAS G12D protein concentration in Figure 2(b). The results indicate the detection limit of this biosensor on KRAS G12D protein detection is $10\ ng/\mu L$ ($0.435\ nM$).

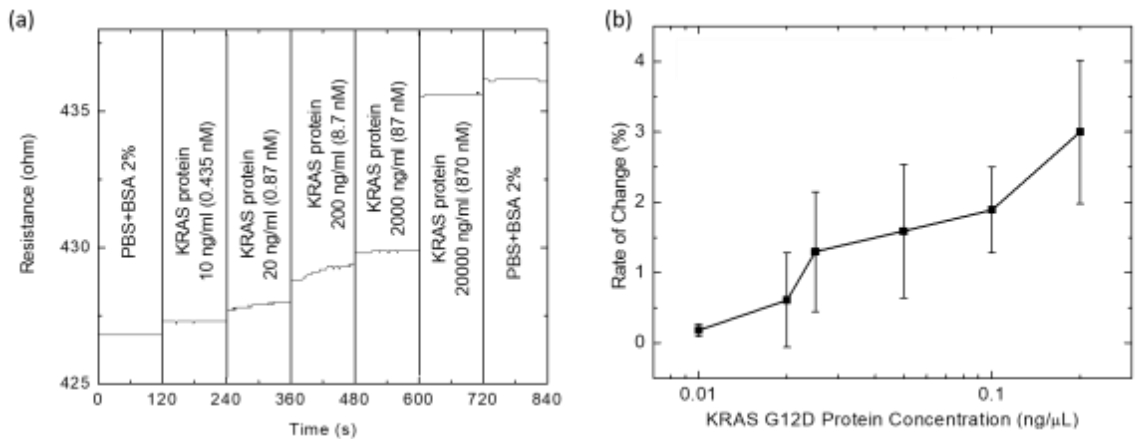

Figure 2. (a) Measured results of a series of KRAS G12D solutions with various concentrations. (b) The rate of change in resistance obtained from multiple measurements of KRAS G12D solutions with various concentrations.

### 3.3 Real-time measurement of clinical specimens

We previously conducted a pancreatic cancer clinical trial with 35 clinical serum specimens using our HEMT biosensing chip[15]. We examined the well-established tumor biomarker of pancreatic carcinoma, carbohydrate antigen 19-9 (CA 19-9), with the threshold of 37 U/mL differentiating patients from healthy people at the detection limit of 7.5 U/mL. While these clinical trials demonstrated an accuracy rate of 94%, the specificity of CA 19-9 for pancreatic cancer is moderate, thus detecting CA 19-9 is suitable for tracking the recovery status of postoperative patients.

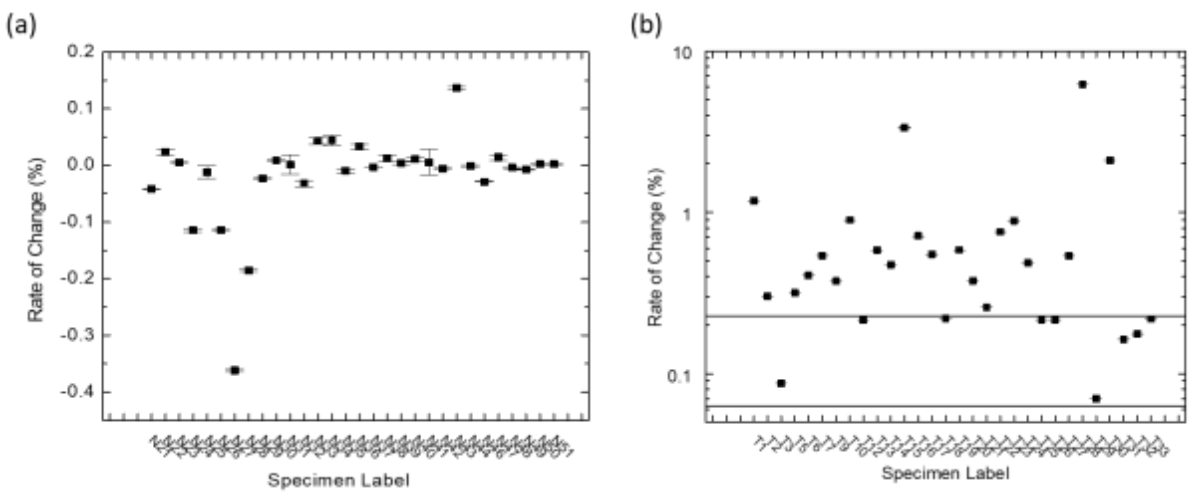

Figure 3. Rate of change of resistance for specimens from (a) 30 healthy donors and (b) 30 PDAC patients. The two horizontal lines in (b) indicate the σ and 3σ above the mean of the rate of change obtained from (a).

KRAS G12D protein, on the other hand, offers higher specificity and appears in earlier stages of pancreatic cancer development[5, 6], therefore is suitable for early detection of pancreatic cancer. As healthy people should not have KRAS G12D mutant protein, no KRAS G12D protein should be detected in serum specimens from healthy people. Clinical specimens from 30 healthy donors and 30 PDAC patients were tested in this report and measured results are provided in Supplementary Information. The rate of change results of 30 healthy donors and 30 pancreatic cancer patients are plotted in Figure 3. The average rate of change from 30 healthy donors is -0.020 with the standard deviation $\sigma = 0.083$. All measured rates of change for 30 PDAC patients exceed one standard deviation above the mean obtained from 30 healthy donors, indicating the effectiveness of the HEMT biosensing chip in discriminating between healthy people and PDAC patients. However, 9 out of 30 patients exhibit the rate of change within three standard deviations above the mean of healthy donors. This suggests that continuous monitoring KRAS G12D mutant protein may be required when the measured rate of change for non-pancreatic cancer patients falls between $\sigma$ and $3\sigma$ above the mean of that measured from healthy people.

## Conclusions

KRAS G12D mutant is a promising biomarker for early detection of pancreatic cancer owing to its high specificity and occurrence in early-stage pancreatic cancer development. We herein performed clinical trials for 30 healthy donors and 30 PDAC patients by detecting KRAS G12D protein from their serum specimens using our HEMT biosensing chip. To the best of our knowledge, this is the first study to detect pancreatic cancer through quantifying KRAS G12D protein. The rate of change in the resistance measured from 30 patients exceeds one standard deviation above the mean obtained from 30 healthy donors, demonstrating the strength of our HEMT biosensing chip. When the rate of change is determined to be between $\sigma$ and $3\sigma$ above the mean of that measured from healthy donors, continuous monitoring the amount of KRAS G12D protein for non-pancreatic cancer patients may be needed to determine whether the person is developing pancreatic cancer. Further studies on detecting KRAS G12D proteins at various stages of pancreatic cancer development are needed to advance early detection of pancreatic cancer.

## Ethics statement

Enrollment and analysis of clinical specimens were conducted per the clinical study protocol approved by the Institutional Review Board of Kaohsiung Medical University Chung-Ho Memorial Hospital (IRB; KMUHIRB-E(I)-20180221) before the initiation of this study. All clinical specimens were informed with consent from patients. All experiments were carried out in accordance with relevant guidelines and regulations as approved by the IRB at Kaohsiung Medical University Chung-Ho Memorial Hospital.

## Acknowledgment

This work was financially supported by the National Science and Technology Council (NSTC), Taiwan, under grants NSTC (112-2112-M-110-005-MY3) for Sheng-Ting Hung and NSTC (114-2112-M-003-015-MY2) for Ching-Wen Chang. Additional support was provided by the joint research program between National Sun Yat-sen University (NSYSU) and Kaohsiung Medical University (KMU) (113KN006). Cheng-Hsuan Chan gratefully acknowledges financial assistance from the Undergraduate Student Research Program funded by NSTC (NSTC (112-2813-C-110-078-B)).

## References

(1) Cancer of the Pancreas—Cancer Stat Facts. Available online: https://seer.cancer.gov/statfacts/html/pancreas.html (accessed on April 3, 2025).
(2) Waters, A. M.; Der, C. J. KRAS: The Critical Driver and Therapeutic Target for Pancreatic Cancer. *Csh Perspect Med* **2018**, *8* (9). DOI: ARTN a031435

10.1101/cshperspect.a031435.
(3) Lee, J. K.; Sivakumar, S.; Schrock, A. B.; Madison, R.; Fabrizio, D.; Gjoerup, O.; Ross, J. S.; Frampton, G. M.; Napalkov, P.; Montesion, M.; et al. Comprehensive pan-cancer genomic landscape of altered cancers and real-world outcomes in solid tumors. *Npj Precis Oncol* **2022**, *6* (1). DOI: ARTN 91

10.1038/s41698-022-00334-z.
(4) Guo, J. L.; Xie, K. P.; Zheng, S. J. Molecular Biomarkers of Pancreatic Intraepithelial Neoplasia and Their Implications in Early Diagnosis and Therapeutic Intervention of Pancreatic Cancer. *Int J Biol Sci* **2016**, *12* (3), 292-301. DOI: 10.7150/ijbs.14995.
(5) Linehan, A.; O'Reilly, M.; McDermott, R.; O'Kane, G. M. Targeting KRAS mutations in pancreatic cancer: opportunities for future strategies. *Front Med-Lausanne* **2024**, *11*. DOI: ARTN 1369136

10.3389/fmed.2024.1369136.
(6) Bannoura, S. F.; Khan, H. Y.; Azmi, A. S. KRAS G12D targeted therapies for pancreatic cancer: Has the fortress been conquered? *Front Oncol* **2022**, *12*. DOI: ARTN 1013902

10.3389/fonc.2022.1013902.
(7) Dong, L. H.; Wang, S. J.; Fu, B. Q.; Wang, J. Evaluation of droplet digital PCR and next generation sequencing for characterizing DNA reference material for mutation detection. *Scientific Reports* **2018**, *8*. DOI: ARTN 9650

10.1038/s41598-018-27368-3.
(8) Orue, A.; Rieber, M. Optimized Multiplex Detection of 7 KRAS Mutations by Taqman Allele-Specific qPCR. *Plos One* **2016**, *11* (9). DOI: ARTN e0163070

10.1371/journal.pone.0163070.
(9) Li, X. Z.; Yang, T. Y.; Li, C. S.; Song, Y. T.; Lou, H.; Guan, D. G.; Jin, L. L. Surface Enhanced Raman Spectroscopy (SERS) for the Multiplex Detection of Braf, Kras, and Pik3ca Mutations in Plasma of Colorectal Cancer Patients. *Theranostics* **2018**, *8* (6), 1678-1689. DOI: 10.7150/thno.22502.
(10) Genco, E.; Modena, F.; Sarcina, L.; Björkström, K.; Brunetti, C.; Caironi, M.; Caputo, M.; Demartis, V. M.; Di Franco, C.; Frusconi, G.; et al. A Single-Molecule Bioelectronic Portable Array for Early Diagnosis of Pancreatic Cancer Precursors. *Advanced Materials* **2023**, *35* (42). DOI: ARTN 2304102

10.1002/adma.202304102.
(11) Dai, E. Y.; Han, L.; Liu, J.; Xie, Y. C.; Kroemer, G.; Klionsky, D. J.; Zeh, H. J.; Kang, R.; Wang, J.; Tang, D. L. Autophagy-dependent ferroptosis drives tumor-associated macrophage polarization via release and uptake of oncogenic KRAS protein. *Autophagy* **2020**, *16* (11), 2069-2083. DOI: 10.1080/15548627.2020.1714209.
(12) Chen, X.; Zeh, H. J.; Kang, R.; Kroemer, G.; Tang, D. L. Cell death in pancreatic cancer: from pathogenesis to therapy. *Nat Rev Gastro Hepat* **2021**, *18* (11), 804-823. DOI: 10.1038/s41575-021-00486-6.
(13) Li, C. B.; Chen, X. H.; Wang, Z. H. Review of the AlGaN/GaN High-Electron-Mobility Transistor-Based Biosensors: Structure, Mechanisms, and Applications. *Micromachines-Basel* **2024**, *15* (3). DOI: ARTN 330

10.3390/mi15030330.
(14) Chang, C. W.; Chen, P. H.; Wang, S. H.; Hsu, S. Y.; Hsu, W. T.; Tsai, C. C.; Wadekar, P. V.; Puttaswamy, S.; Cheng, K. H.; Hsieh, S.; et al. Fast detection of tumor marker CA 19-9 using AlGaN/GaN high electron mobility transistors. *Sensor Actuat B-Chem* **2018**, *267*, 191-197. DOI: 10.1016/j.snb.2018.04.008.
(15) Chang, C. W.; Hsu, Y. N.; Yang, Y. H.; Chang, S. W.; Zhang, J. W.; Chen, Q. Y.; Hung, S. T.; Cheng, K. H.; Hsieh, S. C.; Wang, H. Y.; et al. A Fast CA 19-9 Screening for Pancreatic Cancer Clinical Trials Utilizing AlGaN/GaN High Electron Mobility Transistors. *Ieee Sens J* **2024**, *24* (20), 31754-31762. DOI: 10.1109/Jsen.2024.3445715.
(16) Kim, J. E.; Lee, K. T.; Lee, J. K.; Paik, S. W.; Rhee, J. C.; Choi, K. W. Clinical usefulness of carbohydrate antigen 19-9 as a screening test for pancreatic cancer in an asymptomatic population. *J Gastroen Hepatol* **2004**, *19* (2), 182-186. DOI: DOI 10.1111/j.1440-1746.2004.03219.x.
(17) Chang, C. Y.; Huang, S. P.; Chiu, H. M.; Lee, Y. C.; Chen, M. F.; Lin, J. T. Low efficacy of serum levels of CA 19-9 in prediction of malignant diseases in asymptomatic population in Taiwan. *Hepato-Gastroenterol* **2006**, *53* (67), 1-4.
(18) Kim, B. J.; Lee, K. T.; Moon, T. G.; Kang, P.; Lee, J. K.; Kim, J. J.; Rhee, J. C. How do we interpret an elevated carbohydrate antigen 19-9 level in asymptomatic subjects? *Digest Liver Dis* **2009**, *41* (5), 364-369. DOI: 10.1016/j.dld.2008.12.094.
(19) Tong, Y. L.; Song, Z. Y.; Zhu, W. H. Study of an elevated carbohydrate antigen 19-9 concentration in a large health check-up cohort in China. *Clin Chem Lab Med* **2013**, *51* (7), 1459-1466. DOI: 10.1515/cclm-2012-0458.
(20) Kim, S.; Park, B. K.; Seo, J. H.; Choi, J.; Choi, J. W.; Lee, C. K.; Chung, J. B.; Park, Y.; Kim, D. W. Carbohydrate antigen 19-9 elevation without evidence of malignant or pancreatobiliary diseases. *Scientific Reports* **2020**, *10* (1). DOI: ARTN 8820

10.1038/s41598-020-65720-8.
(21) Tsai, S. S.; George, B.; Wittmann, D.; Ritch, P. S.; Krepline, A. N.; Aldakkak, M.; Barnes, C. A.; Christians, K. K.; Dua, K.; Griffin, M.; et al. Importance of Normalization of CA19-9 Levels Following Neoadjuvant Therapy in Patients With Localized Pancreatic Cancer. *Ann Surg* **2020**, *271* (4), 740-747. DOI: 10.1097/Sla.0000000000003049.
(22) Hornbeck, P. V.; Kornhauser, J. M.; Tkachev, S.; Zhang, B.; Skrzypek, E.; Murray, B.; Latham, V.; Sullivan, M. PhosphoSitePlus: a comprehensive resource for investigating the structure and function of experimentally determined post-translational modifications in man and mouse. *Nucleic Acids Res* **2012**, *40* (D1), D261-D270. DOI: 10.1093/nar/gkr1122.
(23) Baur, B.; Steinhoff, G.; Hernando, J.; Purrucker, O.; Tanaka, M.; Nickel, B.; Stutzmann, M.; Eickhoff, M. Chemical functionalization of GaN and AlN surfaces. *Appl Phys Lett* **2005**, *87* (26). DOI: Artn 263901

10.1063/1.2150280.